\documentclass[sigconf]{acmart}

\usepackage{booktabs}
\usepackage{tabularx}       % better tables
\usepackage[super]{nth}     % for easier ordinal number
\usepackage{enumitem}                % better lists
\usepackage{multicol}
\usepackage{appendix}

\copyrightyear{2021}
\acmYear{2021}
\setcopyright{cc40}

\acmConference[AIES '21] {Proceedings of the 2021 AAAI/ACM Conference on AI, Ethics, and Society}{May 19--21, 2021}{Virtual Event, USA.}
\acmBooktitle{Proceedings of the 2021 AAAI/ACM Conference on AI, Ethics, and Society (AIES '21), May 19--21, 2021, Virtual Event, USA}
\acmPrice{}
\acmISBN{978-1-4503-8473-5/21/05}
\acmDOI{10.1145/3461702.3462605}

\definecolor{darkgray}{RGB}{100,100,100}
\definecolor{darkplum}{RGB}{50,10,50}
\definecolor{Mahogany}{RGB}{103,10,10}
\definecolor{RedInstruct}{RGB}{180,45,45}

\newcommand\aff[1]{\textcolor{darkplum}{{\emph{--#1}}}}

  \newcommand\q[1]{\textcolor{Mahogany}{\small{\textbf{}}}}

\newcommand\gray[1]{\textcolor{darkgray}{#1}}

\newenvironment{lq2}
{ \begin{itemize}[leftmargin = 2.0em, rightmargin=1.0em, label={}]
    \fontsize{8.3pt}{8.9pt}\selectfont
    \setlength{\itemsep}{3pt}
    \setlength{\parskip}{3pt}
    \setlength{\parsep}{3pt}     }
{ \end{itemize}                  }

\newenvironment{lq1}
{ \begin{itemize}[leftmargin = 0em, label={}]
    \fontsize{8pt}{8.6pt}\selectfont
    \setlength{\itemsep}{2pt}
    \setlength{\parskip}{2pt}
    \setlength{\parsep}{2pt}       }
{ \end{itemize}                    }

 \def\Exciting/{{\fontfamily{lmss}\selectfont\textbf{Exciting}}}  \def\Useful/{{\fontfamily{lmss}\selectfont\textbf{Useful}}}
 \def\Worrying/{{\fontfamily{lmss}\selectfont\textbf{Worrying}}}
 \def\Futuristic/{{\fontfamily{lmss}\selectfont\textbf{Futuristic}}}
 \def\None/{{\fontfamily{lmss}\selectfont\textbf{None}}}

\newcommand\question[1]{\bigskip\noindent\textcolor{darkgray}{\large{\textbf{#1}}}\newline}
\newcommand\sample[1]{\noindent\{\textcolor{RedInstruct}{#1}\}\newline}
\newcommand\questiontext[1]{\noindent{#1}\newline}
\newcommand\openend{\newline\indent\{Open-end\}}

\begin{document}
\fancyhead{}

\title{Exciting, Useful, Worrying, Futuristic:\\ Public Perception of Artificial Intelligence in 8 Countries}

\author{Patrick Gage Kelley}
\affiliation{\institution{Google}\country{USA}}
\email{patrickgage@acm.org}

\author{Yongwei Yang}
\affiliation{\institution{Google}\country{USA}}
\email{yongwei@google.com}

\author{Courtney Heldreth}
\affiliation{\institution{Google}\country{USA}}
\email{cheldreth@google.com}

\author{Christopher Moessner}
\affiliation{\institution{Ipsos}\country{USA}}
\email{christopher.moessner@ipsos.com}

\author{Aaron Sedley}
\affiliation{\institution{Google}\country{USA}}
\email{asedley@google.com}

\author{Andreas Kramm}
\affiliation{\institution{Google}\country{Switzerland}}
\email{akramm@google.com}

\author{David T. Newman}
\affiliation{\institution{Google}\country{USA}}
\email{dtnewman@google.com}

\author{Allison Woodruff}
\affiliation{\institution{Google}\country{USA}}
\email{woodruff@acm.org}



\renewcommand{\shortauthors}{Kelley, et al.}

\begin{abstract}
As the influence and use of artificial intelligence (AI) have grown and its transformative potential has become more apparent, many questions have been raised regarding the economic, political, social, and ethical implications of its use. Public opinion plays an important role in these discussions, influencing product adoption, commercial development, research funding, and regulation. In this paper we present results of an in-depth survey of public opinion of artificial intelligence conducted with 10,005 respondents spanning eight countries and six continents. We report widespread perception that AI will have significant impact on society, accompanied by strong support for the responsible development and use of AI, and also characterize the public's sentiment towards AI with four key themes (exciting, useful, worrying, and futuristic) whose prevalence distinguishes response to AI in different countries.
\end{abstract}

\maketitle

\section{Introduction}
As the influence and use of artificial intelligence (AI) have grown and its transformative potential has become more apparent~\cite{horvitz2012interim, stone2016artificial}, many questions have been raised regarding the economic, political, social, and ethical implications of its use~\cite{hawking2014}. The development and application of AI increasingly features in media, academic, industrial, regulatory, and public discussions~\cite{dietterich2015rise, fast2017long, highlevel2019}, with active debate on wide-ranging issues such as the impact of automation on the future of work~\cite{brynjolfsson2014second, raghavan2020, sanchez-mondero2020}, the interaction of AI with human rights issues such as privacy and discrimination~\cite{abrassart2018, barabas2018, buolamwini2018, chancellor2019}, the ethics of autonomous weapons~\cite{scharre2018army, west2018}, and the development and availability of dual-use technologies such as synthetic media that may be used for either benevolent or nefarious purposes~\cite{openAI2019}.

Public opinion plays an important role in these discussions, influencing numerous stakeholders including advocacy groups, funding agencies, regulators, technology companies, and others~\cite{castro2019, cave2018portrayals}. While there have been some explorations of public perception of AI, for example, survey research~\cite{arm2017, blumberg2019, cave2019, ipsos2019, mozilla2019, northeastern2018, west2018, zhang2019artificial}, sentiment analysis~\cite{fast2017long, garvey2019sentiment}, and narrative analysis~\cite{cave2018portrayals}, much of this work has been done in Western, English-speaking contexts. Even in these better studied contexts, much remains to be learned, as both the technology and the public discussion are evolving rapidly. In this paper, we extend previous work by presenting a survey of public perception of AI conducted with 10,005 respondents spanning eight countries and six continents. Our contributions are as follows:

\begin{itemize}[leftmargin = 1.1em]
\item We enrich current understanding of public perception by presenting results of an in-depth survey focused on AI, conducted across a broad range of countries, including several developing countries (encompassing in total: Australia, Canada, the United States (US), South Korea, France, Brazil, India, and Nigeria)
\item We report widespread belief that AI will have significant impact on society, including positive expectations of AI in healthcare, as well as concerns about privacy and job loss
\item We report that a plurality of participants believe that the overall impact of AI could be either positive or negative depending on what happens, and widespread support for responsible development and use of AI
\item We identify four key sentiments (exciting, useful, worrying, and futuristic) whose prevalence distinguishes responses to AI in different countries
\end{itemize}

In the remainder of the paper, we review relevant background, describe our methodology, present and discuss our findings, and conclude.

\section{Background}

Artificial Intelligence (AI) is a broad term with no consensus definition~\cite{edelman2019, fast2017long, stone2016artificial},
and the scope of our inquiry is intended to be similarly broad.
We note that interpretation of the term is further confounded by the ``AI effect'' (the phenomenon that once AI successfully solves a problem and the solution becomes commonplace, it is no longer considered to be AI)~\cite{mccorduck2004machines}, as well as lack of awareness of algorithmic processing in common systems~\cite{eslami2015always, rader2015understanding, warshaw2016intuitions}. 
To aid comparison with our participants' responses,
following~\cite{stone2016artificial}, we share with the reader the following definition provided by Nils J. Nilsson: ``Artificial intelligence is an activity devoted to making machines intelligent, and intelligence is the quality that enables an entity to function appropriately and with foresight in its environment.''~\cite{nils2009}

\subsection{Empirical Studies}
Much of the research on public perception of AI has been survey-based, often conducted in Western, English-speaking countries such as the US and the UK~\cite{blumberg2019, cave2019, edelman2019, zhang2019artificial, northeastern2018} although this has been broadening recently.
AI is often viewed as likely to have a significant impact on the future, with a frequent expectation that its effects will be positive.
In a 2019 Edelman survey in the US, 9 out of 10 respondents assumed that AI will be life-changing and transformational~\cite{edelman2019}.
A Gallup survey conducted in the US in 2018 found that 76\% believed that AI will have a positive impact on their lives~\cite{northeastern2018}; 61\% of respondents had a positive view of AI and robots in a large-scale 2017 survey across Europe on the impact of digitization and automation on daily life~\cite{european2017}; and a 2017 consumer research survey conducted across North America, Europe, and Asia revealed a predominant expectation that society will become better (61\%) rather than worse (22\%) due to increased automation and AI~\cite{arm2017}. A recent Pew Research survey conducted across the Americas, Europe, and Asia showed a somewhat narrower margin (possibly due to shifting public opinion, or alternatively, methodological differences), with (53\%) saying that AI has been mostly good for society versus mostly bad (22\%)~\cite{funk2020}. Considering expected impact in the next 20 years, the 2019 World Risk Poll indicated AI would mostly help (41\%) versus mostly harm (30\% ) people in one's own country, with more favorable impressions in Asia and less favorable impressions in Western countries~\cite{lloyds2020, neudert2020}.

At the same time, AI is neither interpreted as exclusively beneficial nor exclusively disadvantageous, and public response often indicates contradictory emotions. Looking at broad reactions, Blumberg reported that US respondents were equally split between feeling optimistic and informed and feeling fearful and uninformed about AI~\cite{blumberg2019}, while~\cite{arm2017} also revealed both excitement and concern. Relatedly, a 2019 Mozilla survey open to respondents on the Internet gathered continent-level demographic data and revealed varying and mixed emotions at the continent-level~\cite{mozilla2019}. Specific concerns have been expressed regarding social issues, such as AI benefiting the wealthy and harming the poor, fear that AI-enabled deepfakes will erode trust in information, and AI increasing social isolation and reducing human capability~\cite{edelman2019}. In line with these concerns, Zhang and Dafoe found that 82\% of Americans want AI and robots to be carefully managed~\cite{zhang2019artificial}, with 88\% of Europeans expressing similar sentiment~\cite{european2017}. Moreover, 60\% of the general population in the Edelman survey expressed the need for more regulation regarding AI development and deployment~\cite{edelman2019}.

Qualitative work has also explored public perception of algorithmic systems, for example, finding that perception of algorithmic systems can vary substantially by individual factors as well as platform~\cite{devito2017platforms}, and that end users often have fundamental questions or misconceptions about technical details of their operation~\cite{bucher2017algorithmic, eslami2015always, rader2015understanding, ur2012smart, warshaw2016intuitions}.

\subsection{Narratives and Media Sentiment Analysis}
AI is not only heavily discussed in academia, but is also a popular topic in public media~\cite{edelman2019}. 58\% of the respondents in a recent Blumberg survey indicated that they get information about AI from movies, TV, and social media ~\cite{blumberg2019}. In a 2016 CBS news survey, only 19\% indicated not having seen any of several AI movies such as ``The Terminator'' or ``I, Robot''~\cite{cbs2016}. Cave et al. argue that prevalent AI narratives in the English-speaking West share ``a tendency towards utopian or dystopian extremes,'' cautioning that inaccurate narratives could affect technological advancement and regulation~\cite{cave2018portrayals}, with similar points raised in~\cite{horvitz2012interim, stone2016artificial, you2015}. Cave et al. surveyed UK respondents regarding their responses to eight dominant narratives about AI, reporting that the strong majority elicited more concern than excitement~\cite{cave2019}. At the same time, sentiment analysis of newspaper articles from the New York Times and associated content found that, in general, AI has had consistently more optimistic than pessimistic coverage over time~\cite{fast2017long}, and did not support the hypothesis that news media coverage of AI is negative~\cite{garvey2019sentiment}.

\subsection{National Considerations}
A number of countries have established national strategies to promote the use and development of AI, which vary by country and may influence public perception~\cite{dutton2018}.\footnote{See also \url{https://futureoflife.org/national-international-ai-strategies/}} 
The importance of studying local context is also illustrated by analysis of country-specific opportunities and challenges for AI, e.g.~\cite{kalyanakrishnan2018opportunities}.
Further, researchers have called for better integration of developing country considerations in the discussion and development of AI~\cite{sambasivan2019toward}.

\section{Methodology}
In order to better understand public perception of AI, 
we partnered with Ipsos, a global market research firm, 
to conduct a survey of 10,005 respondents in eight countries in July 2019. Methodologically, this work falls in the genre of public opinion polling, as described below.

\subsection{Instrument Development and Translation}
To develop concepts and questions, we consulted experts at our institutions, reviewed published work, drew on our own previous unpublished research, and conducted an initial pilot survey in June 2018 with 1300 respondents drawn from a panel of the general online population in the US. Many questions in the final instrument~\footnote{For the questions used in the instrument see: \href{https://arxiv.org/abs/2001.00081}{arXiv:2001.00081}} were written uniquely for this survey while others were modified from or replicate other questions in the literature or the canon of public opinion surveys. In order to more accurately reflect real-world settings, we did not define AI, and left interpretation of the term to the respondents.\footnote{We note that in our pilot, we had two versions of the survey (one that defined AI and one that did not) and responses to subsequent questions were similar regardless of whether a definition had been provided.} We included primarily closed-form questions as well as a few open-ended questions for free responses. We also included standard demographic questions such as age, gender, education, income, region, and urbanicity. 
% The final instrument included several dozen questions on a range of topics related to artificial intelligence; in this paper we focus on select questions related to the following research questions:
The final instrument included several dozen questions on a range of topics related to artificial intelligence.
In this paper we focus on select questions related to the research objectives below.

Recognizing that public response and adoption are informed by intertwined cognitive and emotional factors~\cite{homburg2006, lin2004}, we explore the following research objectives (the first being cognitive, the second being emotional, and the third considering how they vary by country):

\begin{itemize}[label={--}]
\item What social impact do respondents anticipate AI will have?
\item How do respondents feel about AI?
\item How do the issues above vary by country?
\end{itemize}

After completing the instrument in English, we engaged cApStAn, a linguistic quality assurance agency with expertise in survey translation. We made several improvements based on their insights to minimize terminology that would be difficult to translate. In consultation with cApStAn, we also developed a translation style guide and question-by-question translation guidelines for each target language, for example, specifying translations of key terms and standard scales for Likert questions in each language to ensure consistency (see Table~\ref{tab:demographics} for the languages we offered).

Our market research partner's in-country translation teams and/or third party vendors then translated the full instrument to all target languages while referring to the style guides and guidelines. When the instrument was fully translated, it was provided to us for final review, and the translations were revised through an iterative process before final sign-off and deployment. After the survey was complete, a professional translation vendor provided verbatim translations for all non-English responses; these verbatim translations are used in example quotes in this paper.

\begin{table*}\centering
% \ra{1.3}

\begin{tabularx}{\textwidth}{@{}l XXXXXXXX@{}}

\emph{Country} & \bf AUS & \bf CA & \bf US & \bf KR & \bf FR & \bf BR & \bf IN & \bf NG \\

\midrule

\emph{HDI Rank}     & \nth{8} & \nth{16} & \nth{17} & \nth{23} & \nth{26} & \nth{84} & \nth{131} & \nth{161} \\

\midrule

\emph{Languages}   & \small{English} & \small{English,}  & \small{English} & \small{Korean} & \small{French} & \small{Brazilian}    & \small{English,} & \small{English} \\
\emph{offered}      &         & \small{French}      &       &        &        & \small{Portuguese}   &          \small{Hindi} &  \\

\midrule
\emph{Weighting}    
                    & \small age, gender, \newline education, \newline region
                    & \small age, gender, \newline education, \newline region
                    & \small age, gender, \newline education, \newline region, race
                    & \small age, gender, \newline education, \newline region
                    & \small age, gender, \newline education, \newline region
                    & \small age, gender, \newline education, \newline region
                    & \small age, gender, \newline education
                    & \small age, gender, \newline education \\
      
\specialrule{0.1em}{4pt}{3pt}

%\small{\emph{\# of}}       & & & & & & & \\
\small{\emph{Respondents}} & 1000 & 1500 & 1501 & 1000 & 1001 & 1503 & 1500 & 1000 \\

\midrule

%% WEIGHTED AGE
%\emph{\smaller Weight Age}   & \small{$M=42$}    & \small{$M=44$}    & \small{$M=40$}    & \small{$M=43$}    & \small{$M=40$}    & \small{$M=34$}    & \small{$M= 30$}   & \small{$M= 30$}  \\
%             & \scriptsize{$SD=16.5$} & \scriptsize{$SD=15.7$} & \scriptsize{$SD=13.1$} & \scriptsize{$SD=15.5$} & \scriptsize{$SD=13.1$} & \scriptsize{$SD=12.1$} & \scriptsize{$SD=15.2$} & \scriptsize{$SD=9.3$} \\
 
%% REAL AGE
 \emph{Age}   & \small{$M=43$}    & \small{$M=44$}    & \small{$M=44$}    & \small{$M=40$}    & \small{$M=43$}    & \small{$M=34$}    & \small{$M=30$}    & \small{$M=31$}  \\
             & \scriptsize{$SD=15.3$} & \scriptsize{$SD=16.0$} & \scriptsize{$SD=17.5$} & \scriptsize{$SD=12.4$} & \scriptsize{$SD=15.4$} & \scriptsize{$SD=12.3$} & \scriptsize{$SD=8.9$} & \scriptsize{$SD=9.0$} \\       
\midrule
        
%\emph{\smaller Weight Gend}  & \small{49\%}\scriptsize{ male}   & 
%                 \small{50\%}\scriptsize{ male}   & 
%                 \small{50\%}\scriptsize{ male}   & 
%                 \small{50\%}\scriptsize{ male}   & 
%                 \small{51\%}\scriptsize{ male}   & 
%                 \small{49\%}\scriptsize{ male}   & 
%                 \small{61\%}\scriptsize{ male}   & 
%                 \small{65\%}\scriptsize{ male}   \\
                 
%               & \small{51\%}\scriptsize{ female}  
%               & \small{50\%}\scriptsize{ female}    
%               & \small{50\%}\scriptsize{ female}    
%               & \small{50\%}\scriptsize{ female}    
%               & \small{49\%}\scriptsize{ female}    
%               & \small{51\%}\scriptsize{ female}    
%               & \small{39\%}\scriptsize{ female}    
%               & \small{35\%}\scriptsize{ female}    \\

\emph{Gender}  & \small{49\%}\scriptsize{ male}   & 
                 \small{47\%}\scriptsize{ male}   & 
                 \small{49\%}\scriptsize{ male}   & 
                 \small{52\%}\scriptsize{ male}   & 
                 \small{50\%}\scriptsize{ male}   & 
                 \small{49\%}\scriptsize{ male}   & 
                 \small{70\%}\scriptsize{ male}   & 
                 \small{63\%}\scriptsize{ male}   \\
                 
               & \small{51\%}\scriptsize{ female}  
               & \small{53\%}\scriptsize{ female}    
               & \small{51\%}\scriptsize{ female}    
               & \small{48\%}\scriptsize{ female}    
               & \small{50\%}\scriptsize{ female}    
               & \small{51\%}\scriptsize{ female}    
               & \small{30\%}\scriptsize{ female}    
               & \small{37\%}\scriptsize{ female}    \\
               
\bottomrule

\end{tabularx}

\caption{ Country details, respondent summary and demographics. }
\label{tab:demographics}

\end{table*}

\subsection{Deployment}
We selected a range of countries with different characteristics, such as stage of technological development, nature of the workforce, and varied development indices. The survey was fielded to online panels (groups of respondents who have agreed to participate in surveys over a period of time) representative of the online population in each country. Consistent with the best panels available for online market research, such panels tend to be broadly representative of the general population in countries with high access to technology, but less representative of the general population in countries with more limited access to technology; for example, in developing countries they tend to skew urban. Respondents were recruited using stratified sampling (a method of recruiting specific numbers of participants within demographic subgroups), with hard quotas on age and gender in each country. A summary of countries and demographics is provided in Table~\ref{tab:demographics}. 

The median survey length was 23 minutes across all completions, including those who said they had never heard of AI in an early screening question and received a much shorter variant of the survey. All respondents received incentives in a point system or cash at an industry-standard amount for their market.

\subsection{Data Processing and Analysis}

\subsubsection{Quality Checks}
The market research firm conducted quantitative and qualitative checks to remove low quality responses on an ongoing basis until the quota was reached in each country. Example grounds for removal included being identified as a bot, speeding (answering substantially more quickly than the median time), or providing nonsensical or profane responses to open-ended questions. Overall we removed 7.3\% of responses for quality. After data collection was complete, standard procedures were followed to apply a modest weighting adjustment to each respondent so that the samples in each country are more representative~\cite{biemer2008weighting}. The variables considered in weighting appear in Table~\ref{tab:demographics}.

\subsubsection{Coding of Open-Ended Responses}\label{Coding_Section}
We reviewed the open-ended responses from the pilot to identify emergent themes~\cite{beyer1997contextual} and develop an initial codebook for all questions, then iterated as we reviewed responses from all countries to refine it as necessary.
The open-ended responses were coded in the native language by our market research partner's dedicated coding team (English and French) and one of their third party coding vendors (all other languages).
As described in McDonald et al., a variety of different approaches may be employed to improve the reliability of qualitative analysis~\cite{mcdonald2019}.
In our case, following best practices in public opinion research for coding against multiple languages, we used professional coders, followed an iterative process to continuously improve the codes, and performed a series of hierarchical quality checks.
While coders were specialized by language, they worked together to ensure consistency, sharing notes in specialized coding software. Both we and our market research partner performed multiple levels of quality checks on the resulting coding, randomly sampling from all responses in each country as well as checking all instances of select codes. 

For the open-ended question regarding the feelings or emotions the respondent associated with AI, we began by following the process described above; the resulting codebook for this question encompassed 92 codes (e.g. `Useful,' `Skeptical,' `AI takes over') and specified that multiple codes could be assigned per response. After these codes were assigned and we reviewed the open-ended verbatim responses in detail, four groups of codes emerged from the data as common and semantically distinct (\Exciting/, \Useful/, \Worrying/, and \Futuristic/) -- for example, the \Useful/ group encompassed codes such as `Useful,' `Helpful,' `Productivity,' etc. 
We assigned each of the 92 codes to exactly one of these four groups or Other accordingly. Other encompassed answers that were inarticulate, classified as unable to be coded, mentions of technology without any sentiment (e.g. ``computer" or ``technology"), and a long tail of other opinions on AI (for example ``curiosity" or ``surprise"). Based on the codes that each response had been assigned, each response was considered to be part of those group(s) -- for example, if a response had been assigned the code `Helpful' and the code `Concern,' that response was part of the sentiment groups \Useful/ and \Worrying/. A response that received \textit{only} codes labeled Other appears in \None/.

When we report statistical results for open-ended questions below (for example, the frequency of a code for a given question), these results apply only to the specific question. When we present illustrative quotes, we often draw responses from across any of our four open-ended questions, as relevant responses and similar coding often applied across them.

\subsubsection{Analysis}\label{Analysis_Section}
We used an inductive approach which involved exploring emerging themes and common patterns in the data~\cite{hinkin1998brief}. 
%Quantitative data analyses were performed using SPSS Statistics. All statistical tests assumed $p<.05$ as a significant level. 
When participants qualified to take the survey, they were required to provide an answer to each question, though many questions offered an option for the respondent to indicate they did not know the answer. All respondents were shown three questions and select additional questions were filtered if participants had never heard of AI. In this report, we designate participants who had at least heard of AI as `AI-aware.'

% HDI
As the impact and use of AI expands worldwide, how people learn about, interact with, and use AI varies. People from developed countries (i.e. countries that are more industrialized and have higher per capita incomes, which include Australia, Canada, the US, South Korea, and France) have different needs than people from developing countries (i.e. countries that are less industrialized and have lower per capita incomes, which include Brazil, India, and Nigeria), and this shapes how AI is perceived, adopted, and normalized globally~\cite{un2014, sambasivan2019toward}. Therefore, we anticipated that there might be meaningful differences in AI perceptions associated with development level. We include the Human Development Index (HDI) Rank in Table~\ref{tab:demographics}.\footnote{We show HDI ranks from the 2020 Human Development Report \url{http://hdr.undp.org/en/content/developing-regions}, which uses HDI values from 2019, aligning with the dates of our survey deployment.} 

%We also explored demographic correlations, such as the impact of age, gender, urbanicity, education, income, and employment. Broadly, the differences between individual countries were much more salient, so in this paper we focus on developed-developing and country-level differences rather than demographic correlations.

%We used a chi-square test of homogeneity to analyze participants' attitudes about the long-term impact of AI across jobs,
% quality of life, 
%healthcare, personal relationships, and privacy.
%Post hoc analysis involved pairwise comparisons using the z-test of two proportions with a Bonferroni correction. Given our a priori hypothesis that there would be differences in AI perceptions in developed and developing markets, we ran a t-test when appropriate to compare attitudinal differences in the impact of AI and technology on participants' lives.

% STATS CUT -- also because this isn't clear anyway
% After the sentiment groups were created, we computed correlations by re-coding our covariates of interest and looking at their associations with the sentiment groups. All correlations were pooled and weighted by country-wise sample size.

\subsection{Limitations}

We note several limitations of our methodology that should be considered when interpreting this work. First, it carries with it the standard issues attendant with survey methodology, such as the risk of respondents misunderstanding questions, poor quality translation, or respondents satisficing~\cite{holbrook2003telephone} or plagiarizing open-ended responses. We have worked to minimize these risks through piloting, use of open-ended questions in conjunction with closed-form questions, translation style guides and review, and data quality checks. We also note that panels in India are well-known in the industry to be disproportionately likely to have a social desirability response bias (as defined in~\cite{holbrook2003telephone}), so optimism in the Indian responses should be considered in that context. Second, online panels are not representative of the general population. While we have used a high standard of currently available online panels, we caveat our findings as not representative of the general population, particularly in Brazil, India, and Nigeria. Third, while members of the research team and/or market research partner team have experience conducting research in all markets studied, members of the team reside in Western countries. We have worked to minimize the risk of misinterpretation by collaboration and discussion with in-country partner teams but recognize that our interpretations may lack context or nuance that would have been more readily available to local residents.

 %  %  %  %
 %        %
 %        %
 %  %  %  %

\section{Findings}

In this section, we first provide contextualizing information about how respondents describe AI, and how they report learning about it. We then turn to respondents' expectations of the impact of AI on society. We conclude this section by exploring respondents' sentiment regarding AI.

   %  Examples how to use the tables:

   % In Table~\ref{tab:sentimentgroups} we see sentiment groups.
   % In Table~\ref{tab:devspeed} we see speed up / slow down.
   % In Table~\ref{tab:forsociety} we see benefit for society.

\subsection{Description and Exposure}

We begin with brief context on what respondents understand AI to be. In response to an open-ended question in which we asked respondents to describe AI, respondents across countries often mentioned concepts like computers or robots that think independently, learn, or perform human tasks.~\q{U1-U4}

\begin{lq2}
\item a machine that can think for itself~\aff{Australia}\footnote{We use verbatim responses throughout (in some cases translated) and do not correct typographic or grammatical errors.}
\item Extremely calculative robot~\aff{Nigeria}
\item Artificial Intelligence, to my understanding, is the programming of computers in order for them to learn using different experiences and through lots of different examples.~\aff{Canada}
\item It is a robot or device that is made to be able to preform task as good as if not better than a human~\aff{US}
\end{lq2}

Respondents' comments sometimes indicated partial knowledge, associated AI with other technologies such as the Internet or the Internet of Things (IoT), described AI as something non-technological, or indicated that they did not know.

\begin{lq2}
\item Something I do not understand, but which appears to be a major technological revolution.~\aff{Brazil} 
\item Only thing that I can think of is chess playing programs. However, I'm sure there are more applications in use - technology companies must be using AI somehow.~\aff{Australia} 
\item AI is used in many ways which a ordinary person like me won't even be able to think off.~\aff{India}
% \item I think it has to do with technology~\aff{Nigeria}
% \item it's when the internet sort of anticipates your needs~\aff{Nigeria}
\item AI is a type of technology that makes all your gadget or that connects all your tech stuff all together to make life more easier\\~\aff{Nigeria} 
\end{lq2}

Respondents reported learning about AI in many places, the top five being: social media (45\%), TV reports and commentaries (42\%), movies or TV shows (40\%), magazine or online articles (32\%), and family and friends (31\%).~\q{Q6} 
% The highest rated channel anywhere was social media in India (61\%; second place was social media in Nigeria at 60\%).  
While not in the top five, personal experience using products that have AI technology also drives awareness (13\%).~\q{Q6} The examples below illustrate media influence:~\q{U1-U4}

\begin{lq2}
\item A recent bad example with the 737 Max aircraft that made decisions on their own, resulting in two accidents with many victims.~\aff{Brazil}
\item I am intrigued by this idea. Computers being able to write a screen play in 15 minutes is amazing. I am a writer and just heard a pod cast about this.~\aff{US}
\item Does no one watch movies, read, or anything to do with science fiction!!! It ALWAYS ends badly... there is just no good outcome, that I can see (for now at least), to an actual, fully fledged, AI.\\~\aff{US} 
\item I have recently heard in news and everywhere AI is now the latest invention the man can acheive~\aff{India} 
\end{lq2}

    \subsection{Widespread Expectation That AI Will Change The World}

In this section, we present respondents' expectations of the impacts of AI on society. We begin by discussing respondents' strong expectation that AI will have significant impact, and then discuss some of the specific domains in which they expect that impact to occur. We then discuss the valence of their expectations, whether positive, negative, or ambivalent. We close this section by drawing together these ideas with respondents' strong desire for responsible innovation and care in the development and deployment of AI.

\begin{table*}\centering
% \ra{1.3}

\begin{tabular}{@{}l@{\hskip 1cm}rrrrrrrr@{}}

\emph{Country} & \bf AU & \bf CA & \bf US & \bf KR & \bf FR & \bf BR & \bf IN & \bf NG \\

\midrule
Respondents ($n$) \gray{\emph{AI-aware}} & 946 & 1424 & 1406 & 995 & 970 & 1481 & 1472 & 967
\\

\specialrule{0.00em}{7pt}{1pt}

%\multicolumn{9}{l}{\emph{In general, should the development of Artificial Intelligence (AI)...?}}\\

% \midrule
% Speed up & 
% 21\% & 23\% & 20\% & 36\% & 30\% & 46\% & 73\% & 75\%
% \\
% Stay the same & 
% 35\% & 37\% & 38\% & 28\% & 47\% & 41\% & 14\% & 16\%
% \\
% Slow down & 
% 19\% & 17\% & 18\% & 15\% & 12\% &  7\% &  7\% &  5\%
% \\
% Stop & 
%  8\% &  6\% &  7\% &  7\% &  2\% &  1\% &  1\% &  1\%
% \\
% Don't know & 
% 18\% & 17\% & 18\% & 13\% &  8\% &  5\% &  4\% &  4\% \\
% \bottomrule

% \specialrule{0.00em}{7pt}{1pt}

\multicolumn{9}{l}{\emph{Overall, in the long term, Artificial Intelligence (AI) will be...}}\\

\midrule
Mostly good for society & 
18\% &  20\% & 	21\% & 	23\% & 	18\% & 	38\% & 	51\% & 	37\% 
\\
Mostly bad for society & 
14\% &  15\% & 	17\% & 	8\% & 	 14\% & 	 7\% & 	 8\% & 	 4\%  
\\
Either good or bad for society, depending on what happens & 
43\% &  39\% & 	40\% & 	60\% & 	42\% & 	41\% & 	26\% & 	48\%  
\\
Good and bad in roughly even amounts & 
14\% &  17\% & 	13\% & 	6\% & 	 13\% & 	10\% & 	 12\% & 	9\%  
\\
Won't have much effect on society & 
1\% & 	 1\% & 	 1\% & 	 1\% & 	 2\% & 	 1\% & 	 1\% & 	 1\% 
\\
Don't know & 
9\% & 	 8\% & 	 8\% & 	2\% & 	 11\% & 	 4\% & 	 2\% & 	 2\%  
\\
\bottomrule

\end{tabular}

\caption{ Public opinion regarding the long-term impact on society from respondents who reported being aware of AI. }
%\caption{ Public opinion regarding rate of development of AI (top) and the long-term impact on society (bottom). }
\label{tab:speedandimpact}

\end{table*}

Many of our respondents believe that AI will be transformative. Across all eight countries, only a tiny number (1-2\%, $M = 1\%$) indicate that AI ``won't have much effect on society'' in the long term (see Table~\ref{tab:speedandimpact}).~\q{Q17} Similar sentiment can be observed in the open-ended responses:~\q{U1-U4}

\begin{lq2}
\item AI will revolutionise the way we live in our future.~\aff{India}
\item New and improved tech that will change our lives for good\\~\aff{Nigeria}
\item we are entering a new era. Very modern~\aff{Canada}
\item Machines taking over humans!! :) on a serious note, A.I. is making things possible we thought were not possible a few years ago. Computers recognise faces and fingerprints of humans. Machines carry out so many things to assist humans. Everywhere we look there are examples of artificial intelligence around us.\\~\aff{Australia}
\item It makes me think about how this is going to shape our future and I feel excited by that.~\aff{Australia}
\end{lq2}

We see this reflected in specific aspects of life as well, for example, with low levels of people reporting that AI will have ``no change'' on 
%quality of life (10-27\%, $M = 18\%$),~\q{Q9}
measures such as healthcare (9-27\%, $M = 19\%$),\q{Q11} 
privacy (6-25\%, $M = 15\%$),\q{Q14}
and jobs (7-21\%, $M = 15\%$) in the future.\footnote{These ranges are the spread across the eight countries, and $M$ is the response across our entire sample. For example, for privacy: 6\% of respondents thought there would be no change in Nigeria, 25\% thought there would be no change in France, and 15\% across all respondents.}\q{Q8}
Across all countries surveyed, AI was seen as promising for healthcare.
Most respondents believe AI will lead to ``better healthcare'' in the future ($M = 56\%$ ``better'' versus $M = 12\%$ ``worse'' across all countries),\q{Q11} with Nigeria the highest with 78\% (positive-to-negative ratio = 19.5) expecting improvement.
This optimism resonates with media narratives encouraging AI adoption in healthcare efforts.\footnote{\url{https://borgenproject.org/ai-in-african-healthcare-revolutionizing-the-industry}}\textsuperscript{,}\footnote{\url{https://allafrica.com/stories/201907040526.html}}

\begin{lq2}
\item Progress, I know it will impact positively especially in the areas of health care.~\aff{Nigeria}
\item I believe it has great opportunities for advancement in medical technology. It has moderate use in banking. It has the ability to provide useful information instantaneously. It will greatly improve the quality of mundane tasks.~\aff{Canada}
\end{lq2}

Expectations of AI-related change are not always positive, with concerns about privacy, job loss, and harm to personal relationships in all countries.\q{Q11,Q8,Q13}
Respondents saw privacy concerns as a likely downside of AI, with a plurality of respondents in all countries except for India (50-64\% in other countries; 31\% in India) believing they will have less privacy in the future because of AI ($M = 52\%$ ``less'' versus $M = 23\%$ ``more'' across all countries).
% [cut developed-developing] While still a minority view, nearly 4-in-10 respondents in developing countries (39\%, $N=1519$) held the belief that AI would lead to \textbf{more} privacy in the future, compared to 1-in-10 in developed countries (12\%, $N=664$) ($\chi^{2}(1)=983.92, p<.001$), reinforcing the optimism that many in the developing countries have for AI.
~\q{Q14} 

\begin{lq2}
\item A new frontier. Very exciting and scary at the same time. Lots to gain but will personal privacy be the price?~\aff{Australia}
\item optimistic that it will enhance peoples lives and bring about breakthroughs in many fields but also skeptical that people will lose their jobs and there will be an invasion of privacy~\aff{Canada}
\item A trending mobile app that undresses people. It violates privacy rules~\aff{Nigeria}
\item IT DICTATED MY WEIGHT AND HEIGHT IN PUBLIC.\\~\aff{India}
\item Ads that show up on computers after visiting websites is one thing, but ads that show up after just talking about something makes me think my phone is listening in on my conversations\\~\aff{US}
\end{lq2}

Respondents expected that AI will heavily impact the number of jobs available in the future. Across all countries surveyed, many respondents (39-76\%, $M=53\%$) expected that AI will contribute to ``more jobs lost'' in the future, while far fewer expected it to lead to ``more jobs created'' (7-43\%, $M=22\%$).~\q{Q8}
% [cut developed-developing] Respondents in developing countries were more optimistic, less pessimistic, and more certain about the future impact of AI on their job numbers. Concerns about job loss were heightened among respondents from developed countries, who reported more often that AI will lead to ``more jobs lost'' (58\%, $N = 3318$) in the long term compared to those from developing countries (46\%, $N = 1805$; $\chi^{2}(1) = 128.60, p<.001$).
The open-ended data further emphasizes such concerns, illustrating the perception that AI may replace humans or make them less necessary in the workforce, as well as the association of robots in particular with job loss (due to their ability to perform human tasks), and in rare cases personal experiences related to job loss:~\q{U1-U4}

\begin{lq2}
\item I feel that it has taken away jobs~\aff{US}
\item A highly computerised potentially dangerous job stealing system of machinery operation~\aff{Australia}
\item New technologies. Convenience in life. Reduction in jobs.\\~\aff{South Korea}
\item Am happy about it but am still sceptical about it. This is because it might probably put some persons out of work~\aff{Nigeria}
\item Unemployment comes to my mind when I hear the phrase Artificial Intelligence(AI).~\aff{India}
\end{lq2}

Respondents also anticipate that AI will impact personal relationships in the future. South Koreans expressed the highest concern that AI will weaken personal relationships, with 61\% of respondents saying so (34-50\% ``weaker'' across the other countries, all: $M=46\%$, versus 9-45\% ``stronger'', $M=22\%$), resonating with expectations expressed in the open-ended responses that AI and robots will replace humans.

\begin{lq2}
\item fear that during my lifetime I will be interacting more with AI than live humans~\aff{US}
\item It helps the future by making things easier, but diminishes employment and human contact.~\aff{France}
\item Makes life convenient and can replace humans~\aff{South Korea}
\end{lq2}

% \item It can make life for people more convenient and do things that we can't do. But it's still a device (AI) created by men, so seeing that the world is turning less human isn't so nice to see.~\aff{South Korea}

Beyond these specific expectations for sector-based change in healthcare, jobs, privacy, and personal relationships,
across all our respondents 29\% expected AI would be mostly good for society overall.
% HDI
There were clear differences between countries, and by considering HDI rank, we see developing countries taking a more optimistic view of its future effects. Respondents in Brazil, India, and Nigeria reported AI would be mostly good for society (37-51\%) while in Australia, Canada, the US, South Korea, and France that range was lower (18-23\%).
%, $M = 20\%$) ($\chi^{2}(1, N = 2821) = 557.04, p < .001$).~\q{Q17} 
% \begin{lq2}
% \item Good for our future and obviously next step of human evolution.~\aff{India}
% \item It is beneficial for all the mankind~\aff{India}
% \item Can uplift the world and ridden humans from problems~\aff{India}
% \item Using of machine to solve human problems in almost all sphere of life.~\aff{Nigeria}
% \item This is something that can make lives better, we can do this~\aff{Nigeria}
% \end{lq2}
Nonetheless, ambivalence was evident in all eight countries. Across all countries, 41\% of respondents believe that in the long term AI will be either good or bad for society, depending on what happens, and 12\% believe it will be good and bad in roughly even amounts.~\q{Q17} Respondents sometimes shared mixed emotions in open-ended responses as well:~\q{U1-U4}

\begin{lq2}
\item It is a wonderful and terrifying concept that is inevitable.\\~\aff{Australia}
% mixed
\item A little excitement and a little terror~\aff{US}
% mixed
\item The future of our world in a way that represents both progress and destruction~\aff{Canada}
% mixed
\end{lq2}

Respondents further highlighted that the effects of AI could be either positive or negative, or both positive and negative:

\begin{lq2}
\item Mixture of amazement at the potential of this technology and concern about possible pitfalls. Could be the start of something amazing or the beginning of the end (a la Terminator).\\~\aff{Australia}
% mixed, could go either way
\item A mixture of knowledge and fear. I know that it will help or is already helping in several important areas, but there is always that fear that one of these AIs will become too autonomous and turn against us.~\aff{Brazil}
% mixed, good and bad
\item Life will be much more enjoyable, but I fear that we'd lost what makes us human. Robots will replace humans in various fields, but there are positive sides as well, a pet robot being one of them.~\aff{South Korea}
% good and bad
\item Artificial intelligence is something most people will come to depend on in a few decades. It will make life easier at the same time make people lose their jobs. But one I'm certain of is that AI is here to stay for good.~\aff{Nigeria}
% good and bad
\item It can help us a great deal in the future if it is used for the good of humanity, but we also run the risk of all this software generating major chaos!~\aff{Brazil}
% could go either way
\item Unsure about the net value - has lots of positives but also there are some very legitimate concerns.~\aff{Canada}
% good and bad
\item It's exciting to think about the things that could come about with AI that would make our lives easier and safer, but also scary of course, who knows how it will truly effect society~\aff{US}
% mixed, could go either way
\end{lq2}

This potential for AI to have variable effects was reflected in all eight countries in response to questions about the importance of large technology companies following responsible innovation practices. 82\% of respondents thought it was very or extremely important for these companies to ``Carefully weigh the pros and cons before releasing new technologies’’ and similarly 81\% thought they should ``Follow a responsible process to develop products that use AI.’’ These strong levels of support illustrate the high value respondents place on responsible development and use of AI.~\q{Responsible Development} Open-ended responses also revealed respondents' expectation that the effects of AI depend on whether it is used responsibly.

\begin{lq2}
\item Artificial intelligence worries me a bit because if it's not used well it can be dangerous, it has no conscience or ethics, but I acknowledge that it is an amazing tool.~\aff{France}
\item Depends in the context. Customer service ai? Great. Ai drones - very bad~\aff{Australia}
\item A bit excited because it makes job quite easy but again its scary if it the technology goes wrong like someone using it for evil purposes.~\aff{Nigeria}
\item Angry that future concerns or negative impacts aren't ever considered before technology is developed~\aff{Australia}
% \item Path breaking technology which hopefully should be useful for the mankind if used properly.~\aff{India}
\item Artificial Intelligence is very useful for whole human world. But don't use it in a bad way~\aff{India}
\end{lq2}

    \subsection{Sentiment Groups}

\begin{figure*}[!ht]
\centering
  \includegraphics[width=\textwidth]{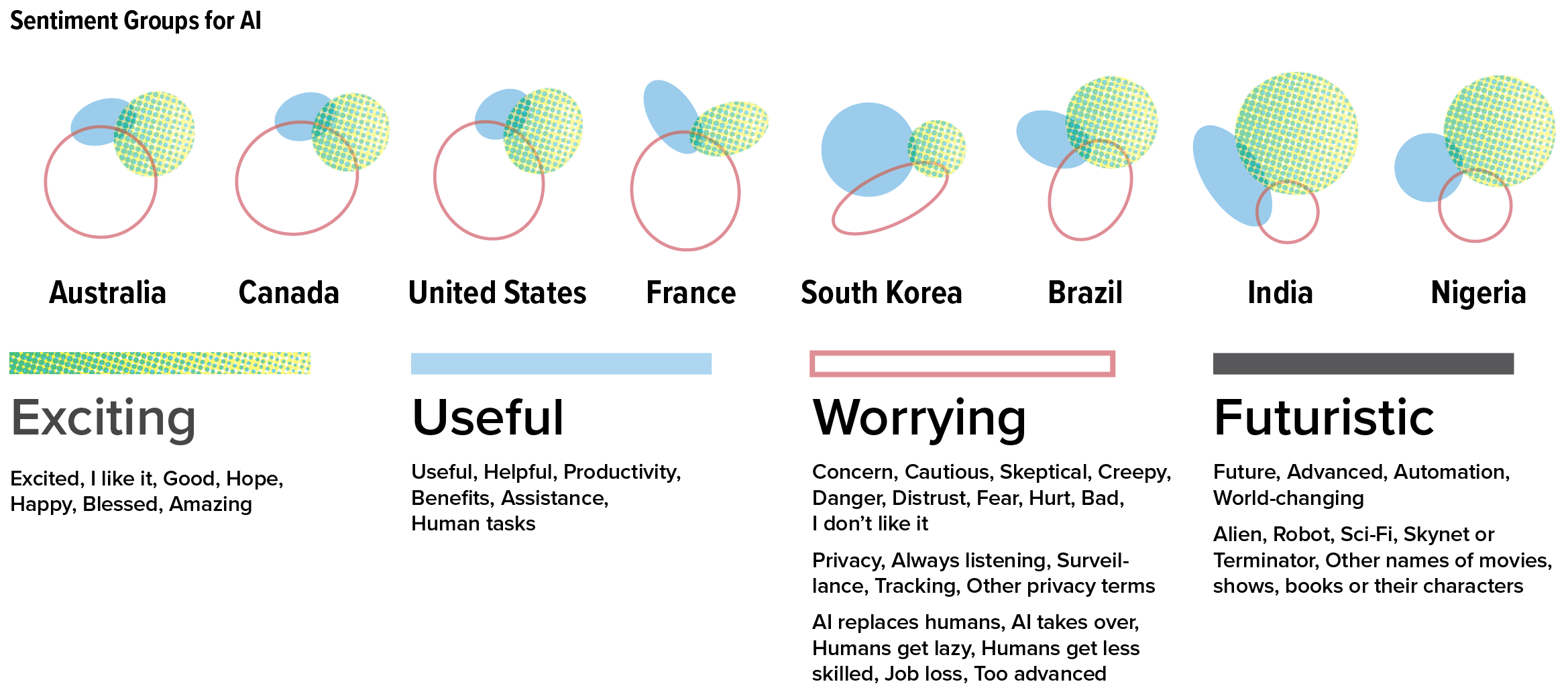}

  \input{tables/groupquotes.tex}
  \vspace{-0.3cm}
  \caption{Description of our four sentiment groups, with the complete list of codes that comprises each, and example responses. While we use the responses to illustrate a particular sentiment, some of them fall in multiple sentiment groups, as sometimes occurred in our data set. At the top of the figure, we represent the overlap between the groups with Venn diagrams, using 3-Venn diagrams which exclude Futuristic for readability. The alert reader may wonder why we use oblong circles; these more accurately represent the area in the overlap. We use the method and tooling described in \protect\cite{micallef2014eulerape}. As throughout the paper we order by HDI, with the exception of South Korea and France, to highlight how similar France is to Australia, Canada, and the United States.}
  \label{fig:venn}
  \vspace{0.2cm}

%\begin{table*}[ht]\centering
% \ra{1.3}

\begin{tabular}{@{}l@{\hskip 1cm}rrrrrrrr@{}}

\emph{Country} & \bf AU & \bf CA & \bf US & \bf KR & \bf FR & \bf BR & \bf IN & \bf NG \\

\midrule
Respondents ($n$) \gray{\emph{All}} & 1000 & 1500 & 1501 & 1000 & 1001 & 1503 & 1500 & 1000 \\

\midrule
\Exciting/ &    17\% & 14\% & 15\% &  6\% &  10\% & 23\% & 36\% & 25\% \\
\Useful/ &       9\% &  9\% &  7\% & 19\% & 11\% & 14\% & 18\% & 11\% \\
\Worrying/ &    31\% & 33\% & 30\% & 14\% & 31\% & 21\% & 9\% &  11\% \\
\Futuristic/ &  22\% & 21\% & 19\% & 38\% & 20\% & 34\% & 24\% & 19\% \\
\None/ &        38\% & 39\% & 42\% & 31\% & 39\% & 25\% & 27\% & 41\% \\

\bottomrule

\end{tabular}
\captionsetup{width=.8\linewidth}
\captionof{table}{
 Percentage of respondents from each country whose open-ended sentiment was coded to be in one of our groups. Respondents can appear in multiple sentiment groups, however a respondent whose answers received only codes not in these groups appears in \None/. }
\label{tab:sentimentgroups}

%\end{table*}

\end{figure*}

Respondent sentiment towards AI was concordant with the results presented in the previous sections, with respondents demonstrating emotions reflective of AI's potential impact. In this subsection we describe the sentiment groups that emerged from our analysis and explore their variation across countries.

Responses to the open-ended question `What feelings or emotions come to mind when you hear the phrase Artificial Intelligence (AI)?'~\q{U1} were assigned to sentiment groups as described in the methodology section.
% Section~\ref{Coding_Section}. 
Multiple assignments were possible, so a response such as ``fear and excited at the same time'' (US Respondent) would be included in \Worrying/ and \Exciting/, but not \Useful/ or \Futuristic/. See Figure~\ref{fig:venn}.

\begin{itemize}

\item \Exciting/ (18.9\%, $N=1888$) -- responses in this group contained positive feelings about AI and often exhibited broad excitement or enthusiasm.

\item \Useful/ (12.2\%, $N=1223$) -- responses in this group expressed the belief that AI will be helpful and assist humans in completing tasks.

\item \Worrying/ (22.7\%, $N=2269$) -- unlike the previous two groups, which each capture a relatively tight set of responses to AI in our open-ended data, this category comprises a wide range of negative emotional responses, predominantly various forms of concern and fear.

\item \Futuristic/ (24.4\%, $N=2439$) -- responses in this group are not necessarily positive or negative towards AI,\footnote{Although \Futuristic/ may not traditionally be seen as a sentiment, when we asked participants to describe their feelings or emotions about AI, they organically responded at very high rates (19\% to 38\% across the countries).} but rather are included for any mention of the futuristic nature of AI, whether by simply describing AI as advanced; mentioning robots, aliens, or other science-fiction concepts; or by referencing the future directly. 
\end{itemize}

The groups above do not cover all responses. About a third of our sample (34.9\%, $N = 3489$) fell into the \None/ group described in the methodology section.
% \ref{Coding_Section}. 
This 34.9\% of responses were assigned only the 48 codes that we did not include in the groups. See Table~\ref{tab:sentimentgroups}.

We now turn to country-level observations, where we see strikingly different national patterns in response towards AI across the eight countries we studied. Four countries most often find AI \Worrying/; three countries find AI to be predominantly \Exciting/; and only our South Korean respondents are most likely to discuss how \Useful/ AI can be. We visually represent the character of these differences in Figure~\ref{fig:venn}.

% HDI
Consistent with our expectation that developed countries (those most-developed, by HDI rank) would share similarities, the dominant sentiment group in Australia, Canada, and the US was \Worrying/, followed by \Futuristic/ (see Table~\ref{tab:sentimentgroups}).
This resonates with claims that popular press and media narratives in these regions have emphasized potential threats of AI~\cite{cave2018portrayals, fast2017long, horvitz2012interim}.
France shares characteristics with Australia, Canada, and the US, with similar values for the \Useful/, \Worrying/ and \Futuristic/ groups, but with fewer in the \Exciting/ group.

\begin{figure*}[!ht]
\centering
  \includegraphics[width=0.8\textwidth]{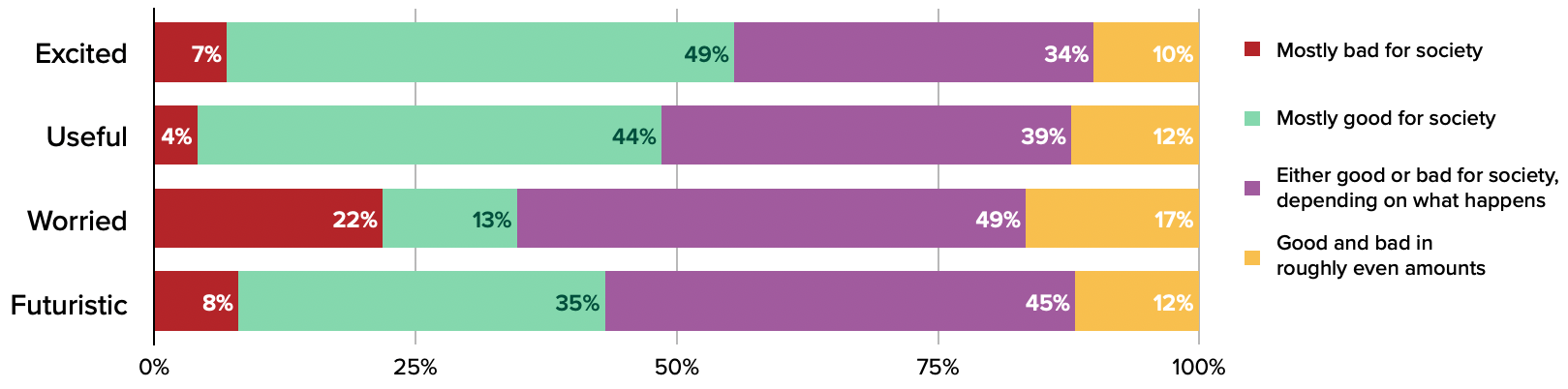}
 \caption{For each sentiment, we show all respondents who were coded as having that sentiment, grouped by their answer to how AI would impact society.}
% \caption{Each respondent that we coded to have expressed one of the four sentiments, is shown above grouped by their answer to how AI would impact society.}
  \label{fig:expsent}
\end{figure*}

South Korea has a unique profile among the countries surveyed, having the largest percentage in the both the \Useful/ (19\%, $N = 192$) and \Futuristic/ (38\%, $N = 379$) sentiment groups. South Korean respondents also had the lowest percentage of \Exciting/ (6\%, $N = 63$ vs 10-37\% in all other countries). 
%Our analysis also reveals that more educated respondents in South Korea were more likely to associate AI with \Useful/ ($r(9924) =.18$, $p < .001$). 
These findings are consistent with South Koreans' high level of exposure to technology: South Korea boasts the world's highest robot density~\cite{ifr2018}, is one of the largest global investors in smart buildings~\cite{ducker2019}, and may be ``at the vanguard of a revolution in AI and big data healthcare''~\cite{nature2019}. Consistent with this, South Korean respondents often mentioned AI assistants and home automation, which may contextualize AI as a more familiar, everyday technology:

\begin{lq2}
\item AI is everywhere from hospitals to homes and cars.\\~\aff{South Korea}
\item Use big data to make daily life more convenient.~\aff{South Korea}
\item Convenient, relaxed daily life~\aff{South Korea}
\item With just the smartphone, I can check the gas, temperature, and the foods in the fridge.~\aff{South Korea}
\item Self-driving car, automated production, convenient daily life\\~\aff{South Korea}
\end{lq2}

% HDI
In the developing countries (those less-developed by HDI rank, i.e., Brazil, India, and Nigeria), \Exciting/ was the dominant sentiment, as seen in Figure~\ref{fig:venn}/Table~\ref{tab:sentimentgroups}. 
Brazil also has a unique profile among the countries surveyed, with higher levels of \Worrying/ and \Futuristic/ relative to the other developing countries. However, Brazilian respondents reported levels of \Exciting/ and \Useful/ similar to those in the other developing countries.
%Across all countries, higher exposure to AI was positively associated with \Exciting/ sentiment ($r(9661) =0.13$, $p< .001$).~\q{Q5}
%Brazil also has a unique profile among the countries surveyed, with higher levels of \Worrying/ (21\%, $N = 319$) and \Futuristic/ (34\%, $N = 504$) relative to other developing countries (\Worrying/ -- Nigeria, 11\%; India, 9\%; \Futuristic/ -- Nigeria, 19\%; India, 24\%). However, Brazilian respondents reported levels of \Exciting/ (23\%, $N = 345$) and \Useful/ (14\%, $N = 211$) similar to those in other developing countries (\Exciting/ -- Nigeria, 25\%; India; 36\%; \Useful/ -- Nigeria, 11\%; India, 18\%).
India was the most enthusiastic about AI, with the highest level for \Exciting/, the second highest level for \Useful/, and the lowest level for \Worrying/ across all countries. 

We now consider the relationship between sentiment towards AI and expected impact on society. In Figure~\ref{fig:expsent}, we show how the respondents we included in each of our four sentiment groups reported believing AI would impact society. For example, of the 1802 respondents who expressed the sentiment \Exciting/, 49\% of them said AI would be mostly good for society, while only 13\% of the 2129 respondents who expressed \Worrying/ thought this. While we do see some overlap between the expectations of AI and our sentiment groups, the sentiments we observed do not fully account for responses regarding societal impact, and vice versa. This provides further support for measuring public opinion of both cognitive and emotional factors regarding AI, as well as laying the groundwork for further investigation of what drives these factors as we discuss further below.

\section{Discussion}
Based on our respondents views of both how they expect AI will transform society and how they feel about AI, we argue for three areas of continued investment: design and ethics resources to help more responsibly build the systems that the public believes will shape the future; additional investment and research into information campaigns to educate about AI, as well as increased design and development of offerings the public values; and further research into the media narratives that are shaping public expectations and sentiment, particularly analysis that explores the nuanced space beyond positive versus negative valence.

\subsection{AI Design and Ethics Guidance}

Our results on public interest in the transformative nature of AI and its responsible use underscore the importance of the
growing body of AI design guidance and ethical toolkits.
Such resources provide practical strategies to responsibly build and deploy systems which the public expects will matter so much to society, and our findings argue for their increased prioritization and application.

AI designers and developers can leverage work on the relationship between AI and HCI (under the names Human-Centered Machine Learning~\cite{lovejoy2017}, Machine Learning UX or MLUX ~\cite{carney2019}, and similar terms), such as Amershi et al.'s synthesis of twenty years of AI design learnings into 18 guidelines for human-AI interaction design~\cite{amershi2019}, the People + AI Guidebook,\footnote{\url{https://pair.withgoogle.com/}} and other resources outlined in Carney's summary~\cite{carney2019}.

Regarding ethical resources, our findings reinforce the importance of designing and developing AI responsibly to benefit society and minimize potential harms~\cite{abrassart2018, highlevel2019}, and of sharing information about those efforts. Recent analyses summarize the rapidly increasing number of principles and guidelines for ethical AI~\cite{fjeld2020,jobin2019}, and tactical support for applying these ideas in practice is available in resources such as the Markkula Center Ethics in Technology Practice Framework and Toolkit,\footnote{\url{https://www.scu.edu/ethics-in-technology-practice/}} the Omidyar Ethical OS Toolkit,\footnote{\url{https://ethicalos.org/}} and the Princeton Dialogues on AI and Ethics Case Studies.\footnote{\url{https://aiethics.princeton.edu/case-studies/}}

\subsection{Interventions and Communications}
Public misperception or unrealistic expectations of AI can lead to unfounded fears or disappointment and disillusionment~\cite{blumberg2019, cave2018portrayals}. 
Our insights into public opinion, particularly in countries where little data has been gathered previously, suggest areas in which the public may benefit from additional information and educational programs.
They also suggest ways in which the design and development of particular technologies may have a favorable impact on public opinion.
For example, future research could explore the conditions facilitating South Korea's unusually strong impression of AI as \Useful/, to gain insight into whether or how this sentiment might resonate elsewhere via communications or technological offerings.
As another example, our findings point to the value of emphasizing AI's application to healthcare in communications as well as product and research investments.
Finally, many respondents were concerned about negative impacts of AI on privacy, reinforcing the value of continued emphasis on designing and developing AI with privacy in mind, concordant with discussion of privacy by design in the EU General Data Protection Regulation (GDPR).\footnote{\url{https://eugdpr.org/}} The privacy discussion continues to evolve quickly, and best practices for AI technologies continue to be actively explored in the academic, legal, and policy communities, offering many opportunities for advances in this area.

\subsection{Narratives and Sentiment Groups}
Our results reflect a number of key dialogues that have appeared in public discussion and the media, for example, that AI offers significant improvements for health; that AI is associated with privacy issues, job loss, and social isolation; and that AI could be either a significant boon or a significant threat to humanity. Other work examining narratives in Western, English-speaking countries has argued that popular portrayals of AI exaggerate this dichotomy~\cite{cave2018portrayals}. Our findings suggest that concern about AI is higher in these countries as well as Australia and France, but is less prominent in South Korea, Brazil, India, and Nigeria.

More broadly, our findings revealed sentiment groups as a distinguishing feature, with respondents in different countries finding AI to be \Exciting/, \Useful/, \Worrying/, and \Futuristic/ to varying degrees. These groups provide one nuanced alternative to understanding people's feelings towards AI, rather than considering their orientation to AI as simply positive or negative. 
Now that these differences between countries have been observed, it would be valuable to learn more about what drives them, e.g. to formally measure and analyze the relationship between media and pop culture narratives in different countries and the presence of these sentiment groups, as well as tracing the relationship and movement of narratives across countries.

Further, it would be useful to explore other factors that likely influence these sentiment groups, such as country culture and economy; presence, awareness, and availability of AI technologies such as customer service chatbots, personal assistants, and more; and personal, formative experiences using AI technology. It would also be worthwhile to explore how sentiment groups affect behavior such as adoption of AI technologies and public opinion on topics such as research funding and regulation.

\section{Conclusions}
We surveyed public opinion of artificial intelligence with 10,005 respondents spanning eight countries across six continents, focusing on issues such as expected impacts of AI, sentiment towards AI, and variation in response by country. We report widespread perception that AI will have significant impact on society but the overall nature of these effects is not yet determined, underscoring the importance of responsible development and use. We identify four groups of sentiment towards AI (\Exciting/, \Useful/, \Worrying/, and \Futuristic/) whose prevalence distinguishes different countries' perception of AI. Our findings suggest opportunities for future work, such as empirical study of the relationship between media narratives and sentiment across countries, and opportunities for interventions and communications regarding the design and development of AI technologies.

\begin{acks}
We thank Elie Bursztein, Ed Chi, Charina Chou, Jen Gennai, Jake Lucchi, Ken Rubinstein, and Kurt Thomas for valuable contributions to this work.
We also thank the cApStAn team for their important contributions to linguistic quality and the Ipsos team for their excellent work fielding the survey.
\end{acks}

\bibliographystyle{ACM-Reference-Format}
\balance
\bibliography{AIsurvey}

\appendix
\section{Questionnaire -- Select Questions}
\textit{Note: Some questions were modified from or replicate other questions in the literature or the canon of public opinion surveys}

\question{Unaided Sentiment}
\sample{Ask All}
\questiontext{What feelings or emotions come to mind when you hear the phrase Artificial Intelligence (AI)?}
\openend

\question{Knowledge}
\sample{Ask All}
\questiontext{How much do you know about Artificial Intelligence (AI)?}
\begin{itemize}
\item{A lot}
\item{A moderate amount}
\item{A little}
\item{Heard of AI, but know nothing about it}
\item{Never heard of AI}
\end{itemize}

\question{Unaided Description}
\sample{Do NOT ask if ``Never heard of AI'' in Knowledge question}
\questiontext{In your own words, please describe Artificial Intelligence (AI).}
\openend

\question{Unaided Examples}
\sample{Do NOT ask if ``Never heard of AI'' in Knowledge question}
\questiontext{Please list some examples of how Artificial Intelligence (AI) is used today.}
\openend

\question{Exposure in Past 12 Months}
\sample{Do NOT ask if ``Never heard of AI'' in Knowledge question}
\questiontext{In the past 12 months, how much have you heard about Artificial Intelligence (AI)?}
\begin{itemize}
\item{A great amount}
\item{A lot}
\item{A moderate amount}
\item{A little bit}
\item{Nothing at all}
\end{itemize}

\question{Source of Learning}
\sample{Do NOT ask if ``Never heard of AI'' in Knowledge question}
\questiontext{In the past 12 months, where have you learned about Artificial Intelligence (AI)?  Please select all that apply.}
\newline\indent{\{Randomize\}}
\begin{itemize}
\item{Popular non-fiction books}
\item{Research journals or papers}
\item{Magazine or online articles}
\item{TV reports or commentaries}
\item{Movies or TV shows}
\item{Social media (e.g., Facebook, Twitter)}
\item{Friends or family}
\item{Meet-up or community groups}
\item{Education or training}
\item{Advertisements}
\item{Using products that have the technology}
\item{Working with the technology myself}
\item{Scientists}
\item{Tech industry leaders}
\item{None of the above \{Anchor, Exclusive\}}
\end{itemize}

% \question{Impact on People's Lives}
% \sample{Do NOT ask if ``Never heard of AI'' in Knowledge question}
% \questiontext{In general, how would you rate the impact of Artificial Intelligence (AI) on people's lives?}
% \begin{itemize}
% \item{Extremely positive}
% \item{Moderately positive}
% \item{Slightly positive}
% \item{Neither positive nor negative}
% \item{Slightly negative}
% \item{Moderately negative}
% \item{Extremely negative}
% \end{itemize}

\question{Future Impacts}
\sample{Do NOT ask if ``Never heard of AI'' in Knowledge question}
\questiontext{In the next ten years, what do you think will happen in [COUNTRY] because of Artificial Intelligence (AI)?}
\newline
\indent{\{Randomize the order of the paired stubs\}}\newline
\indent{\{Show one set at a time\}}\newline
\newline
\indent{More jobs lost}\newline
\indent{No change}\newline
\indent{More jobs created}\newline	
\indent{Don't know}\newline
%\newline
%\indent{Better quality of life}\newline
%\indent{No change}\newline
%\indent{Worse quality of life}\newline	
%\indent{Don't know}\newline
\newline
\indent{Better healthcare}\newline
\indent{No change}\newline
\indent{Worse healthcare}\newline	
\indent{Don't know}\newline
\newline
\indent{Stronger personal relationships}\newline
\indent{No change}\newline
\indent{Weaker personal relationships}\newline	
\indent{Don't know}\newline
\newline
\indent{Less privacy}\newline
\indent{No change}\newline
\indent{More privacy}\newline	
\indent{Don't know}\newline
\newline
\indent{(additional pairs not listed)}

\question{Impact on Society}
\sample{Do NOT ask if ``Never heard of AI'' in Knowledge question}
\questiontext{Overall, in the long term, Artificial Intelligence (AI) will be...}
\newline\indent{\{Rotate scale for top two only for each person\}}
\begin{itemize}
\item{Mostly good for society}
\item{Mostly bad for society}
\item{Either good or bad for society, depending on what happens \{Anchor\}}
\item{Good and bad in roughly even amounts \{Anchor\}}
\item{Won't have much effect on society \{Anchor\}}
\item{Don't know \{Anchor\}}
\end{itemize}

% \question{Rate of Development}
% \sample{Do NOT ask if ``Never heard of AI'' in Knowledge question}
% \questiontext{In general, should the development of Artificial Intelligence (AI) ...?}
% \begin{itemize}
% \item{Speed up}
% \item{Stay the same}
% \item{Slow down}
% \item{Stop}
% \item{Don't know}
% \end{itemize}

\question{Responsible Development}

\questiontext{How important is it for large technology companies to do the following?}
\begin{itemize}
\item{Extremely important}
\item{Very important}
\item{Moderately important}
\item{Slightly important}
\item{Not at all important}
\newline
\end{itemize}

\begin{itemize}
\item[]\{Randomize\}
\item[]{}
\item[]{\sample{Do NOT ask if ``Never heard of AI'' in Knowledge question} Follow a responsible process to develop products that use Artificial Intelligence (AI)}
\item[]{}
\item[]{\sample{Ask All} Carefully weigh the pros and cons before releasing new technologies}
\item[]{}
\item[]{(additional items not listed)}
\end{itemize}

\question{Uncomfortable Experience}
\sample{Do NOT ask if ``Never heard of AI'' in Knowledge question}
\questiontext{Have you ever had an experience with AI-related technology that made you feel uncomfortable?}
\begin{itemize}
\item{Yes}
\item{No}
\item{Not sure}
\end{itemize}

\question{Unaided Description of Uncomfortable Experience}
\sample{Ask if ``Yes'' to Uncomfortable Experience}
\questiontext{What happened, and what was the outcome? Please describe your experience with AI that made you feel uncomfortable.}
\openend

\end{document}